\pdfoutput=1

\documentclass[11pt]{article}

\usepackage{todonotes}
\usepackage[]{acl}

\usepackage{times}
\usepackage{latexsym}
\usepackage{authblk}
\definecolor{LightGray}{gray}{0.9}
\PassOptionsToPackage{dvipsnames}{xcolor}
\usepackage{ragged2e}
\usepackage{booktabs}
\usepackage{multirow}
\usepackage{multicol}
\usepackage[titletoc]{appendix}
\usepackage[hang]{footmisc}
\usepackage{mdframed}
\usepackage{xspace}

\usepackage[T1]{fontenc}

\usepackage[utf8]{inputenc}

\usepackage{microtype}
\usepackage{xspace}

\usepackage{inconsolata}
\usepackage{enumitem}

\newcommand{\semismall}{\fontsize{\semismall@size}{\semismall@baselineskip}\selectfont}

\title{On the Comprehensibility of Multi-structured Financial Documents using LLMs and Pre-processing Tools}

\author{
  Shivani Upadhyay\textsuperscript{1} \quad
  Messiah Ataey\textsuperscript{2} \quad
  Syed Shariyar Murtaza\textsuperscript{2} \quad
  Yifan Nie\textsuperscript{2} \quad
  Jimmy Lin\textsuperscript{1} \\
  \textsuperscript{1}University of Waterloo, Waterloo, Canada \\
  \textsuperscript{2}Manulife, Toronto, Canada \\
  \\
  \texttt{\{sjupadhyay, jimmylin\}@uwaterloo.ca} \\
  \texttt{\{messiah\_ataey, syed\_shariyar\_murtaza, yifan\_nie\}@manulife.com}
}

\begin{document}
\maketitle

\begin{abstract}

The proliferation of complex structured data in hybrid sources, such as PDF documents and web pages, presents unique challenges for current Large Language Models (LLMs) and Multi-modal Large Language Models (MLLMs) in providing accurate answers. 
Despite the recent advancements of MLLMs, they still often falter when interpreting intricately structured information, such as nested tables and multi-dimensional plots, leading to hallucinations and erroneous outputs. 
This paper explores the capabilities of LLMs and MLLMs in understanding and answering questions from complex data structures found in PDF documents by leveraging industrial and open-source tools as part of a pre-processing pipeline. 
Our findings indicate that GPT-$4_o$, a popular MLLM, achieves an accuracy of 56\% on multi-structured documents when fed documents directly, and that integrating pre-processing tools raises the accuracy of LLMs to 61.3\% for GPT-$4_o$ and 76\% for GPT-$4$, and with lower overall cost. The code is publicly available at \url{https://github.com/OGCDS/FinancialQA}.

\end{abstract}

\section{Introduction} 
\label{introduction}
Hybrid data sources such as PDF documents and web pages present information in a variety of structures with complex arrangements. 
The advancements in Multi-modal Large Language Models (MLLMs), including proprietary models like GPT~\cite{gpt4} and Gemini~\cite{gemini}, as well as open-source models such as LlaVA~\cite{liu2023visual, liu2023improved}, LaVIT~\cite{lavit}, and Emu~\cite{emu, emu2}, have led to a significant improvement in these models' abilities to understand heterogeneous data.
However, when it comes to effectively understanding varied structural information sources and providing high-fidelity results, these models still tend to hallucinate and generate incorrect results due to their limited understanding of the input data structure. 
For instance, MLLMs can struggle to interpret tables with complex nesting and certain types of financial plots (e.g., a plot with more than 2 dimensions), limiting their usefulness for question-answering (QA) over complex multi-modal data.

Recent developments in LLMs such as the GPT model family ~\cite{gpt4} and LLaMA ~\cite{Touvron2023LLaMAOA} have shown human-level effectiveness in complex QA tasks. 
One major drawback of such models however is their inability to answer questions about data outside of their training corpus. Retrieval Augmented Generation (RAG) ~\cite{RAGSurvey} is a technique that utilizes LLMs in-context learning abilities by providing relevant information from external sources as part of the prompt, enabling effective QA over such information. 
Despite the ability to retrieve and augment relevant external information, QA over multi-structured document contents such as tables and charts in financial reports still remains a challenge due to their limited reasoning capabilities.

In this paper, industrial and open-source tools are investigated as a way to improve the performance of LLMs for QA over complex multi-structured contents found in PDF documents. 
In particular, Azure Document Intelligence is utilized\footnote{\url{https://azure.microsoft.com/en-us/products/ai-services/ai-document-intelligence}} for the extraction of tabular and textual information, pypdf\footnote{\url{https://pypi.org/project/pypdf/}} for the extraction of images, and the ChartVLM model \cite{chartvlm} for chart comprehension. 
This combination of tools and models ensures a comprehensive extraction and understanding of the common data contents found in PDF documents. These tools are evaluated by comparing them with using the multi-modal GPT-$4_o$ model directly. 
Considering the vast array of tools, models, and platforms available for similar purposes, the study presented in this paper is not intended to be a comprehensive comparison of all possible tools and their combinations. 
Instead, the contribution of this paper is to provide insights into the following research questions:
\begin{itemize}[noitemsep]
\item[\textbf{(a)}] Can MLLMs such as GPT-$4_o$ perform effective QA over multi-structured data contents found in financial documents in the industry when used directly as input?
\item[\textbf{(b)}] Can pre-processing tools help improve the comprehensibility of LLMs and MLLMs for QA over multi-structured documents?
\item[\textbf{(c)}] How can pre-processing tools help save costs when performing QA over multi-structured documents compared to popular end-to-end commercial offerings?

\end{itemize}

Our results show that: (a) GPT-$4_o$, when used with document inputs directly, is only able to achieve 56\% QA accuracy over multi-structured financial PDF documents; (b) utilizing industrial and open-source pre-processing tools improves accuracy to 61.3\% for GPT-$4_o$ and 76\% for GPT-$4$ (both in text-only mode); and (c) the document cost per page is only $\$0.00231$ with pre-processing tools, which is $74.33\%$ cheaper than the cheapest solutions, such as Anthropic's Claude 3 Opus.

\section{Background and Related Works}

\begin{figure}[tbh]
  \centering
   \includegraphics[width=\linewidth]{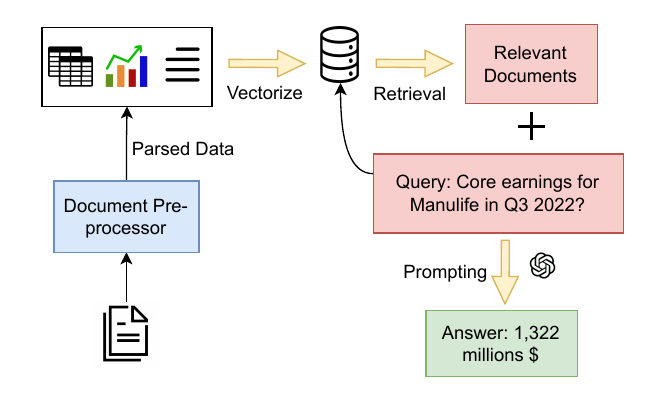}
   \caption{Flowchart of a RAG framework utilizing document pre-processing tools (shown in the blue box). First, multi-structured content is extracted using the document pre-processor. Then, the extracted content is converted to a vector representation and stored in a vector database, which can then be used to retrieve relevant content based on a provided query. The retrieved content is finally augmented with the query in the LLM prompt for relevant in-context generation.}
   \label{fig:flowchart}
\end{figure}

In finance, insurance, and other industries, analysts rely on various documents such as quarterly and annual reports, balance sheets, and financial statements. 
These documents often include complex tables and charts (see Appendix~\ref{doc-ex}), which are essential for making informed decisions.
Due to the large volume of such documents, a generative AI chat application based on an LLM using retrieval augmented generation (RAG) can help analysts find answers to their inquiries faster and with less cognitive overhead. 
Figure~\ref{fig:flowchart} shows a typical RAG-based system along with document pre-processing tools highlighted in blue, wherein information present in textual, graphical, and tabular data formats are extracted. 
The extracted contents are then converted into a vector representation using an embedding model and inserted into an index (e.g., a vector database). 
With such a system, relevant information can then be retrieved from the index at runtime based on similarity to a provided query, which is then augmented with the query in the prompt template to generate relevant responses using an LLM.

\paragraph{Retrieval Augmentation with LLMs:}
RAG has proven to enhance LLMs understanding of external data sources and has shown highly accurate answer generation ~\cite{ lewis2021retrievalaugmentedgenerationknowledgeintensivenlp, nakano2022webgptbrowserassistedquestionansweringhuman, RAGSurvey}. RAG also reduces hallucination as LLMs are provided with relevant information. 
However, even when provided with relevant context, LLMs may struggle to interpret and comprehend certain multi-structured data formats such as complex financial tables and plots.

\paragraph{QA over Tables:}
Analyzing tabular information and answering questions based on them is referred to as Table QA~\cite{TabQAsurvey}. 
Table QA datasets can be categorized into two types: those that use a raw table as direct input for model analysis~\cite{wtq, wikisql}, and those that use a table with additional annotated information~\cite{tatqa, vqaonbd, tabmcqa, finqa}.

\paragraph{QA over Charts:}
Being a crucial method of information representation, Chart QA deals with QA about charts of various types~\cite{ChartQASurvey}. Chart QA has gained a lot of attention as of late, and many datasets have been developed to aid models for enhanced chart comprehension~\cite{chartqa, plotqa, figureqa}.

\paragraph{QA over Documents:}
Many works have been focused on benchmarking and evaluating QA tasks over various types of documents. For example, Visual QA datasets mainly include scanned images of documents~\cite{docvqa, doccvvqa} focusing on text-heavy content,~\cite{narayanan2024aviarytraininglanguageagents, skarlinski2024language, lala2023paperqa} analyzing scientific research and~\cite{qasper} showcases information retrieval from raw documents with \LaTeX \space as input.
The document pre-processing tools investigated in this work specifically focus on extracting multi-structured content found in complex PDF documents, and thus the evaluation dataset used encompasses various data types such as tables and charts.

\section{Methodology for Pre-processing}
In this section, the document pre-processor built using industrial and open-source tools is introduced which transforms a variety of structured information from PDF documents into a simple text format.
Typically, these documents contain charts and nested tables alongside regular text data, which may not be easily interpretable by LLMs. 
This technique transforms a variety of structured information into a simple text format, enabling LLMs to comprehend such complex data structures more easily.

\paragraph{Hybrid Data Sources:}
Financial and business documents contain various data types such as raw text, nested tables, and images. 
Therefore, to gain a complete understanding of a document, it is necessary to extract all of these data types in preferably a standard format. Generally speaking, the raw text contains qualitative information about the topics of a document, and tables and charts convey quantitative information.

\begin{figure}[tbh]
  \centering
  \vspace{-0.35cm}
   \includegraphics[width=\linewidth]{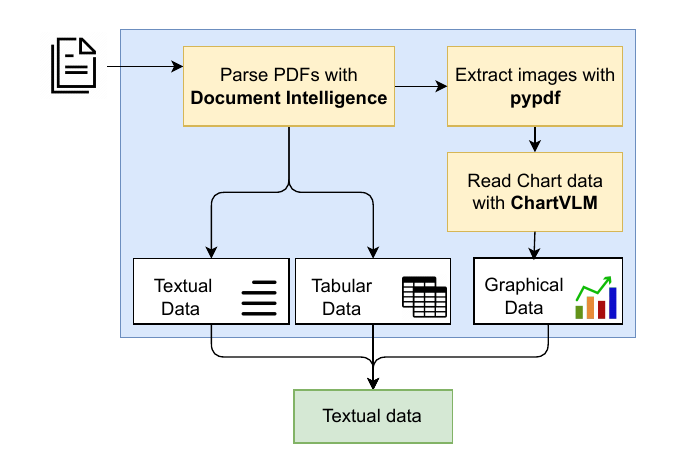}
   \vspace{-0.8cm}
   \caption{The internal workings of the document pre-processor. Azure Document Intelligence is used for the extraction of textual, tabular, and graphical information. pypdf is used for the extraction of images, and ChartVLM is used to convert chart images to their underlying tabular representation.}
   \vspace{-0.5cm}
   \label{fig:parser-flow}
\end{figure}

\subsection{Document Pre-processor}
The pre-processing methodology presented in this work (referred to as a general `pre-processor') handles the extraction of multi-structured data contents (i.e., text, tables, and charts) found in financial PDF documents, as illustrated in Figure~\ref{fig:parser-flow}. 

\subsubsection{Tabular and Textual Data Extraction}
To accurately extract tabular and textual information from documents, Azure Document Intelligence is utilized which combines Optical Character Recognition (OCR) capabilities and advanced deep learning models for precise data extraction. 
Document Intelligence extracts both textual and structural information which can be categorized as either geometric (e.g., text, tables, and figures) or a logical role (e.g., section titles, image captions, page footers). 
Along with role-based data extraction, it also provides the exact coordinates of each extracted object, which could assist in higher-level document analyses. 

Extracted tables are stored in a standard JSON string format, which contains information for each cell in the table including details such as polygon area, row/column-span, and a boolean indicator that signifies whether or not the cell is a column header. 
The column header attribute is particularly useful for capturing key semantic information conveyed by the table. 
In instances where hierarchical column headers are present, the headers are combined in a way to ensure an accurate mapping of the key-value pairs. 
This process allows for a comprehensive and precise extraction and representation of both simple and complex textual and tabular data structures. 
Table~\ref{tab:visual1} shows an example of an extracted table.

\subsubsection{Graphical Data Extraction}
To extract graphical images from PDFs, Azure Document Intelligence is used to extract the bounding box coordinates of an image on a single page, and pypdf to then crop and save the image. In the case of extracting images that are not relevant to a specific task, an optional classifier can be added to evaluate whether an extracted image is of a particular type (e.g., a financial chart) or if it is an incomplete image (e.g., a missing axis of a plot). 
To obtain a textual representation of the extracted financial chart image, a Pix2Struct~\cite{pix2struct}-based ChartVLM model is employed that has been fine-tuned on the ChartX dataset. 
ChartVLM presents an enhanced understanding of various chart types and an inherent ability to preserve the aspect ratio of the input image, which assists with increased robustness in chart data extraction. 
The model converts a chart image into its underlying CSV tabular format which is then converted into the JSON data format as part of the pre-processing pipeline, similar to extracted tables. 
This procedure enables a comprehensive extraction of financial charts from the document. 
A visualization is shown in Table~\ref{tab:visual2}.

An attempt to enhance ChartVLM's interpretation abilities through fine-tuning was carried out, but due to the complexities of real-world chart understanding and limited data availability, the model consistently overfitted and failed to provide any meaningful performance improvements. 
Section~\ref{chart-challenges} provides details about the challenges associated with chart extraction and interpretation.

\section{Experimental Setup}
This section discusses the experimental setup employed to perform QA evaluations. 

\begin{figure*}[!tbh]
        \centering
    \includegraphics[width=1.75\columnwidth]{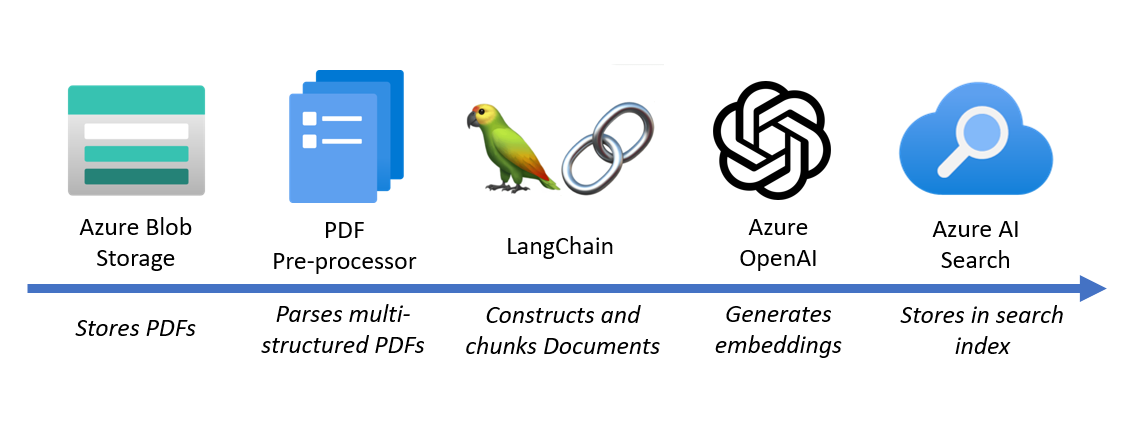}
    \caption{RAG ingestion pipeline used for performing PDF evaluations.}
    \label{fig:ingestion}
\end{figure*}

\subsection{RAG with Document Pre-processing}
In this section, specific details about the RAG with document pre-processing framework are discussed.

\begin{figure}[bth]
\begin{mdframed}[font=\footnotesize, roundcorner=10pt, linewidth=1pt, innerleftmargin=10pt, innerrightmargin=10pt, innertopmargin=5pt, innerbottommargin=5pt]
Comprehend the following context and answer the questions in one line:\\

\{$context$\}\\

Do not add extra information on your own.
\end{mdframed}
\caption{Prompt utilized within RAG framework.}
\label{fig:prompt}
\end{figure}

\subsubsection{Ingestion Pipeline}
The multi-step ingestion pipeline for the RAG framework is shown in  Figure~\ref{fig:ingestion}. 
Initially, PDF documents are collected and optionally stored in an Azure Blob Storage container\footnote{\url{https://learn.microsoft.com/en-us/azure/storage/blobs/storage-blobs-overview}}, or locally on disk. 
These documents are then sent through the document pre-processor to extract multi-structured data contents such as text, tables, and charts. 
The post-processed contents (text), are subsequently truncated into chunks of 600 tokens using LangChain's TokenTextSplitter\footnote{\url{https://js.langchain.com/v0.1/docs/modules/data_connection/document_transformers/token_splitter}} to avoid excessive data sizes, where each chunk corresponds to only a single page of the document if the number of tokens is less than the chunk size. 
Each chunk is then converted to a vector representation using OpenAI's \textit{text-embedding-ada} model. 
The resultant embedding, along with its metadata from the document, is finally uploaded to an Azure AI Search index. 

The metadata consists of contextual information related to the indexed chunk and/or document. 
In the case of a financial document, the metadata might include document-level details such as company name, year, report quarter, etc., and also individual chunk-level details such as page number, section title, bounding-box coordinates, etc. 
Incorporating metadata to the indexed contents is essential as it allows for pre-filtering of the contents based on either user-provided or automatically extracted filters. 
This type of pre-filtering can vastly reduce the volume of indexed contents to be searched against based on the query, thus significantly improving retrieval accuracy.

\subsubsection{LLM Inference}
In the process of query-based retrieval, the three most relevant chunks from the index are retrieved according to their cosine similarity score against the query.\footnote{A value of $k=3$ was empirically chosen as it provided a good trade-off between response accuracy and context size in the experiments.} 
The retrieved data is then optionally processed further and augmented with the query in the prompt to enable in-context generation with an LLM. 
This approach enables the LLM to gain a comprehensive understanding of the provided context and generate accurate responses. 
The prompt template used with the LLM is shown in Figure~\ref{fig:prompt}.

\subsubsection{Dataset} 
For industry-relevant evaluations, a private test set assembled by financial experts was used consisting of 75 question-answer pairs of varying complexity derived from numerous financial documents of Canadian companies, where 53 questions are for tables and 22 for charts. 
Questions of a less complex nature may involve extracting a specific value from a table or simply reading values off a chart. 
Difficult questions on the other hand may necessitate the model to integrate multiple pieces of information to arrive at the final result. 
Appendix~\ref{private-set} provides more details regarding the evaluation set.

\subsection{Individual Components Evaluation}
In this experiment, the tabular and chart components of the document pre-processor are evaluated. 
Tables and charts are extracted from documents and converted to a standard text format using the document pre-processor, which are then passed as input to an LLM, namely GPT-3.5 Turbo and GPT-$4$, to generate answers based on this processed information. 
These experiments do not use a retrieval mechanism since the relevant table or chart is directly provided as input to the LLM. 
Similarly, raw chart images are provided as input to GPT-$4_o$. 

\subsubsection{Dataset} 
The following datasets are used for evaluating LLM table and chart comprehension:

\begin{itemize}[leftmargin=*,noitemsep]
    \item \textbf{VQAonBD dataset:} Tabular images from the dataset (i.e., the validation set of 4535 images) are used as input for the evaluations. The dataset has almost 50 questions partitioned into 5  categories for each image based on their content and format~\cite{vqaonbd}.
    \item \textbf{ComplexChartQA (CCQA) dataset:} 
     A manually-curated chart evaluation dataset consisting of 50 complex chart images along with questions to assess chart comprehension efficacy.
\end{itemize}

\section{Results and Discussion}
To compare with the multi-modal capabilities of GPT-$4_o$, PDF pages were converted into images and subsequently provided as input to GPT-$4_o$. Since GPT-$4_o$ has a maximum input length of 20 images, it was necessary to ensure the answer was contained within the limits of the context size. A performance evaluation of the RAG with document pre-processing framework is also investigated.

\begin{table}[t]
\resizebox{\columnwidth}{!}{%
  \begin{tabular}{l|cclp{1cm}|l|p{1.3cm}} 
    \toprule
    &\textbf{Pre-processing} & \textbf{RAG} & \textbf{LLM} & \textbf{Input mode}&\textbf{Acc.}&\textbf{Cost per call}\\
    \midrule
    \midrule
    1(a) & Y & Y & GPT-$3.5_t$ & T & 60.0 & 0.0003\\
    1(b) & Y & Y & GPT-$4$ & T & \textbf{76.0} & 0.0360\\
    1(c) & Y & Y & GPT-$4_o$ & T & 61.3 & 0.0030\\
    \midrule
    2(a) & - & - & GPT-$4_o$ & I,T & 56.0 & 0.0765\\
    2(b) & Y & Y & GPT-$4_o$ & I & 40.9 & 0.0765\\
    \bottomrule
  \end{tabular}
  }
\caption{End-to-end comparative analysis between GPT-4o and the RAG with pre-processor framework using the private evaluation set. The first section (rows 1*) describes the performance of (text-only) GPT models using the pre-processor, and the second section (rows 2*) of GPT-4o using document pages directly as input.}
\label{tab:complete}
\end{table}

Rows numbered with 1* present a comparison of the  RAG with document pre-processing framework with GPT-3.5 Turbo, GPT-$4$, and GPT-$4_o$ (in text mode (``T'')). Rows numbered with 2* present a comparison of the proficiency of GPT-$4_o$ in analyzing PDF document pages directly, utilizing both image and text (``I, T'') inputs in row 2(a). Row 2(b) presents the results of a multi-modal RAG framework, where raw images are embedded and retrieved from an index, as opposed to plain text. The generated results are subject to assessment by human evaluators.

As shown in row 2(a), the direct application of GPT-$4_o$ to comprehend multi-structured documents yields an accuracy of 56\%. 
When the RAG with document pre-processing framework is used with GPT-$4_o$ in text mode, an improved accuracy of 61.3\% is achieved.
Using GPT-3.5 Turbo with the framework also yields a higher accuracy than GPT-$4_o$ 2(a) at 60.0\%, while using GPT-$4$ significantly improves performance over all models, achieving an accuracy of 76\%. 
This demonstrates the potential of using pre-processing tools in enhancing LLM comprehension of complex multi-structured PDF documents. 
To gain a deeper understanding of context representations and their effectiveness—particularly in the case of complex tables—we conduct an optimal retrieval analysis comparing various representation methods. This analysis enables us to precisely identify the limitations of each representation, independent of retrieval errors.
For further details, please refer to Appendix~\ref{sec:optimal-retrieval}. 

Lastly, the performance of GPT-$4_o$ using a financial chart image-based RAG framework is investigated in row 2(b). In this setting, the images are embedded using a vision-language CLIP model ~\cite{Radford2021LearningTV}, thus enabling retrieval of images using plain text and vice versa. 
Similar to the text-based RAG framework, the top three images are retrieved from the index based on their cosine similarity to the text query. The experiment shows that GPT-$4_o$ performs much worse (40.9\%) on chart QA for the evaluation set considered, likely due to low retriever recall or the retrieval of conflicting information leading to an incorrect response.

\paragraph{Cost Effectiveness:}
An additional advantage of using text-based LLMs is the savings made in API costs. When GPT-$4_o$ is used with both image and text modalities, the average cost per API call can be as high as 0.0765. However, when only the text modality is used with GPT-3.5 Turbo, GPT-$4$, and GPT-$4_o$, assuming $600$ tokens on a typical document page, the cost can be reduced to 0.0003 (i.e., reduced by \textasciitilde200$\times$), 0.0360 (i.e., reduced by \textasciitilde2$\times$), and 0.0030 (i.e., reduced by \textasciitilde20$\times$), respectively.

The pre-processor used in this work is also compared with alternative paid solutions based on the unit cost per page in Table \ref{tab:cost_analysis}, where the same assumption is made as before that a typical document page contains $600$ tokens. 
In the pre-processor, the cost of calling the Azure Document Intelligence and Azure OpenAI's GPT-$4_o$ (in text-mode) API are considered.\footnote{\url{https://azure.microsoft.com/en-us/pricing/details/cognitive-services/openai-service/}} 
The LlamaParse,\footnote{\url{https://docs.cloud.llamaindex.ai/llamaparse/usage\_data}} Vertex AI\footnote{\url{https://cloud.google.com/vertex-ai/generative-ai/pricing\#modality-based-pricing}} and Anthropic\footnote{\url{https://www.anthropic.com/news/claude-3-family}} products are readily available solutions that provide similar functionality.
The vendor prices were converted to a cost per page as shown in Table~\ref{tab:cost_analysis}, where the pre-processor used in this work is shown to be the most cost effective.
\begin{table}
  \centering
    \resizebox{0.3\textwidth}{!}{%
    \begin{tabular}{ccc}
    \toprule
    Solution & Main Model & Cost/Page \\
    \midrule
    \midrule
    LlamaParse & GPT-4$_o$ & \$ 0.03000 \\
    \midrule
    Vertex AI & Gemini-2.0-flash & \$ 0.11610 \\
    \midrule
    Anthropic & Claude 3 Opus & \$ 0.00900 \\
    \midrule
    Our solution & GPT-4$_o$ & \$ 0.00231 \\
    \bottomrule
    \end{tabular}}%
    \caption{Cost comparison with alternative solutions.}
  \label{tab:cost_analysis}
\end{table}%

\begin{table}[tbh]
\resizebox{\columnwidth}{!}{%
  \begin{tabular}{c|l|c c c c c|c} 
    \toprule
    & \multirow{2}{*}{\textbf{Model}} & \multicolumn{5}{c}{\textbf{VQAonBD categories}} & \multirow{2}{*}{\textbf{CCQA}}\\
    & & \textbf{1} & \textbf{2} & \textbf{3} & \textbf{4} & \textbf{5} &\\
    \midrule
    \midrule
    1(a) & pre-processing + GPT-3.5$_t$ & 0.64 & 0.25 & 0.29 & 0.41 & 0.21 & 0.56\\
    1(b) & pre-processing + GPT-$4$ & 0.77 & 0.58 & 0.44 & 0.57 & 0.48 & 0.68 \\
    \midrule
    2 & GPT-$4_o$ (Images) & 0.59 & 0.03 & 0.04 & 0.45 & 0.38 & 0.66 \\
    \bottomrule
  \end{tabular}
  }
  \caption{Optimal retrieval evaluation for pre-processor components.
  }
  \label{tab:table-comp}
\end{table}

\paragraph{Component-Level Analysis:}
Table~\ref{tab:table-comp} showcases the performance of tabular data comprehension for questions of varying levels of difficulty from the VQAonBD dataset (``VGAonBD categories'') and from the ComplexChartQA (``CCQA'') dataset. In this experiment, the question-input pair from the dataset is provided directly to the text-based model for evaluation (row 1*), while for GPT-$4_o$ in image-mode, a raw image of the question-input pair is used (row 2).
For table QA, improvements are observed when using the document pre-processor with GPT-3.5 Turbo and GPT-$4$ as compared to GPT-$4_o$ across all categories. For chart QA, comparable effectiveness is observed between both methods with a slight improvement using the RAG with document pre-processing framework with GPT-$4$.

\section{Discussion}
\paragraph{Challenges with Chart Interpretation.}
\label{chart-challenges}
Interpreting charts is challenging due to the variety of chart types, each requiring a unique approach.
For example, extracting information from bar and line plots requires effectively interpreting their axes and values.
Additionally, even within the same chart type, variations in presentation require the model to recognize and learn from these differences.

Upon manually inspecting the chart results, we find that GPT-$4_o$ demonstrates good interpretation skills when the charts are well labeled and annotated with necessary text values for the datapoints, likely because it can use its OCR abilities to understand the content of the chart.
On the other hand, when the charts do not have such explicit textual content, GPT-$4_o$ often struggles to accurately interpret complex chart elements, leading to a potentially incorrect response. 
Appendix~\ref{sec:gpt-fail} shows an example of this behavior.

ChartVLM demonstrates the capability to interpret chart images without annotations. 
However, its limited comprehension of complex visualizations often results in incomplete or inaccurate interpretations of the entire plot.
Financial documents frequently combine multiple plots within a single image to enhance interpretability.
While this practice is common in real-world scenarios, it falls outside the expected input structure defined by ChartVLM’s internal knowledge, leading to potential misinterpretations.
In some cases, the generated text focuses on only one of the sub-images within a composite chart, overlooking or only partially extracting the remaining components. 
Appendix~\ref{sec:chartvlm-fail} illustrates an example of this behavior.

\paragraph{Challenges with Table Interpretation.}
Interpreting tables becomes increasingly complex when hierarchical structures are involved. The extraction process can be particularly challenging, and the accuracy and reliability of the results heavily depend on precise data extraction.
Many of the errors encountered in the results stem from the loss of hierarchical structure during table parsing. 
When hierarchical JSON data is flattened—as is common with many tables in the dataset—important spatial cues like column alignment and merged cells, which aid human understanding, are lost.
This flattening obscures relationships such as subtotals and groupings, especially when the original layout is not preserved. 
In our case, hierarchical column headers are merged into a single string, while hierarchical row headers are reduced to flat records with empty field values. 
Crafting post-processing heuristics to recover this structure is difficult and not scalable due to the diversity of table formats.

Moreover, in the RAG setting, enriching table chunks with metadata such as section titles, headers, etc. plays a crucial role in resolving ambiguities among similar tables within a document.
This becomes especially important when models mistakenly extract values from incorrect tables due to lexically similar or the same row and/or column headers. Incorporating such contextual metadata aids in both accurate retrieval and generation.

\section{Conclusion}

This paper investigated three primary research questions: whether LLMs can comprehend complex multi-structured PDF contents, the use of pre-processing tools to improve comprehension, and how pre-processing tools can help save costs compared to directly providing document pages to an MLLM. 
The results show that GPT-$4_o$, a widely-used MLLM, answers approximately half of the questions for complex charts and tables. 
When using commercially available and open-source tools for document pre-processing, the response accuracy is improved for all the text-based LLMs considered in the experiments. 
Furthermore, the cost analysis shows considerable cost savings when using the RAG with document pre-processing framework compared to directly using document page inputs with an MLLM.
While such results are promising, it is clear there is still room for improvement of current pre-processing tools and LLMs for comprehending complex multi-structured PDF contents, based on overall accuracy.

Although the source code for the RAG with document pre-processing framework presented in this work is available, including any shareable datasets, the framework is ultimately bounded by licenses and usage agreements of the proprietary GPT models and Azure services.

\bibliography{anthology,custom}

\clearpage

\appendixtitleon
\appendixtitletocon
\begin{appendices}

\section{Private Evaluation Dataset}
\label{private-set}
This section includes details regarding the dataset used for performing the evaluations.
The question set belongs to 4 different financial documents. Table~\ref{tab:private-set} showcases a selection of representative queries extracted from the dataset.
For the LLM to provide an accurate response to the query, it is essential for it to grasp the central concept inherent in the query and potentially employ multiple pieces of reference information to arrive to the answer. Low difficulty questions may involve just reading off values from a chart or table, whereas difficult question may require to perform arithmetic operations and/or multiple pieces of information to be retrieved as context.

\begin{table}[bth]
\resizebox{\columnwidth}{!}{%
  \begin{tabular}{p{4.5cm}|cc}
    \toprule
    \textbf{Query} & \textbf{Difficulty Level} & \textbf{\# of References}\\
    \midrule
    \midrule

    Which asset category has the highest percentage of total invested assets? & Low & 1 \\
    \midrule
    What is the quarterly core ROE change from 2022 to 2023 for Manulife? & Medium & 3\\
    \midrule
    What is the new Q1/23E Core / Underlying YoY in MFC? & High & 5\\
    \bottomrule
  \end{tabular}
  }
\caption{Example queries from the private evaluation dataset. Each query is assigned a difficulty level based loosely on how much "reasoning" is required for accurate QA.}
\vspace{-0.4cm}
\label{tab:private-set}
\end{table}

\section{Comparison of Context Formats for Table}
\label{sec:optimal-retrieval}
\begin{table}[t]
\resizebox{\columnwidth}{!}{%
  \begin{tabular}{cclcl} 
    \toprule
    \textbf{Parser} & \textbf{RAG} & \textbf{LLM} & \textbf{Context type} & \textbf{Accuracy}\\
    \midrule
    \midrule
    Y & - & GPT-$3.5_t$ & JSON & 75.4 \\
    Y & - & GPT-$4$ & JSON & 100.0 \\
    Y & - & GPT-$4_o$ & JSON & 96.2 \\
    \midrule
    Y & - & GPT-$3.5_t$ & Dataframe & 84.9 \\
    Y & - & GPT-$4$ & Dataframe & 96.2 \\
    Y & - & GPT-$4_o$ & Dataframe & 92.4 \\
    \bottomrule
  \end{tabular}
  }
\caption{Direct input analysis with the private evaluation set (tables). Here, the relevant extracted information is passed directly as context to an LLM. The context type field represents the specific format considered for evaluation.}
\label{tab:without-rag}
\end{table}
In this section, a comparative study between the utilization of JSON and Dataframe representation of tables is investigated.
This evaluation assumes an ``optimal retrieval'' scenario, where the correct input is provided directly as input and thus just focuses on comparing the two representation formats. 
As indicated in Table~\ref{tab:without-rag}, both methods exhibited comparable performance.
This comparison allows us to select the best context representation method for the complex hierarchical tables present in the documents.
However, it is noteworthy that for GPT-$4$ and GPT-$4_o$, the use of JSON demonstrated superior efficacy.
Consequently, JSON was chosen as the preferred representation for tabular and chart information.\footnote{Given the relatively lower complexity of contextual representation in charts, a direct comparison wasn't needed. Instead, we adopted a unified JSON-based approach for representing both structures.}

\section{Visualizations}
\subsection{Document Example}
\label{doc-ex}
\begin{figure*}[bth]
        \centering
    \includegraphics[width=1.75\columnwidth]{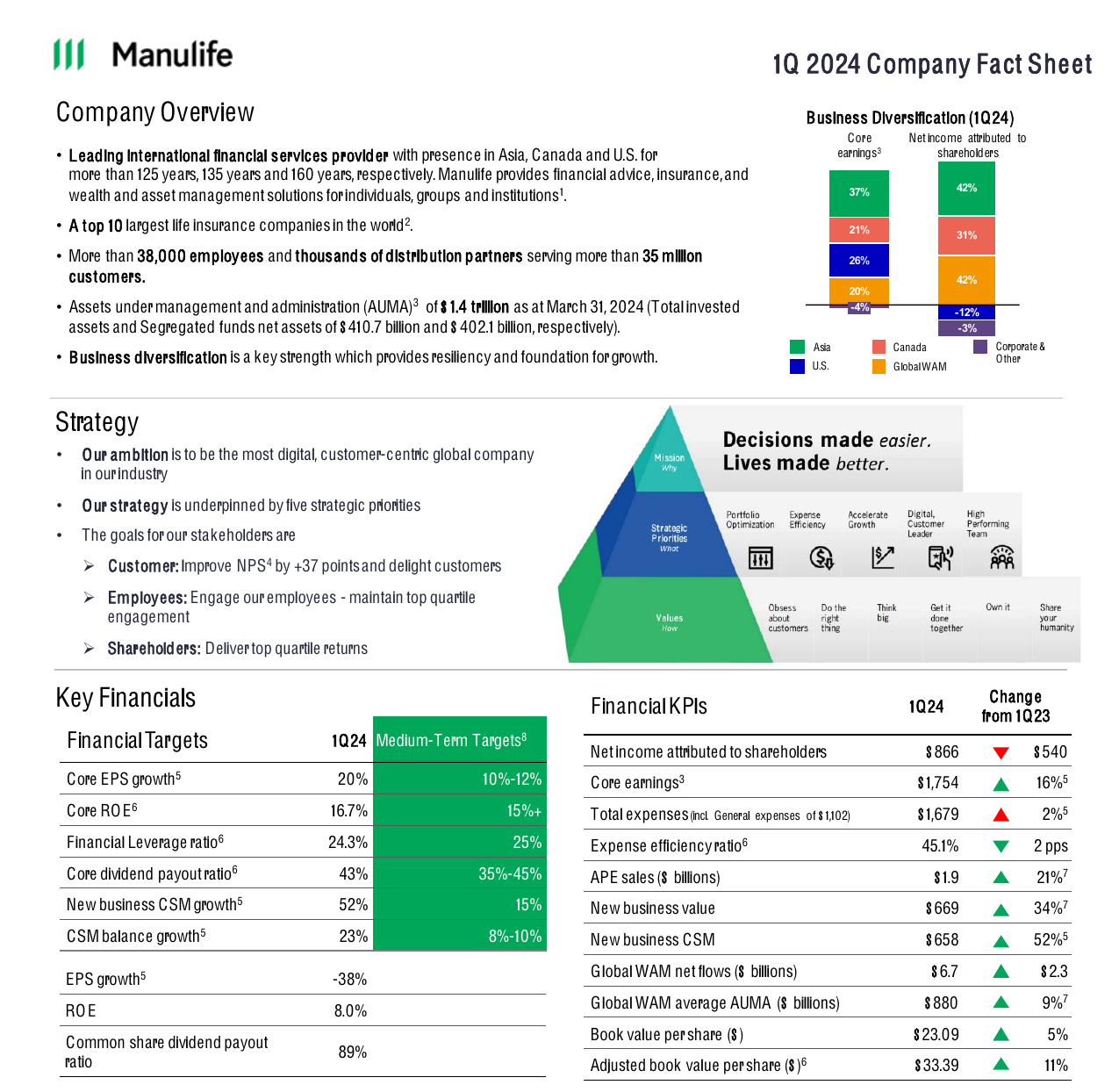}
    \caption{Financial document example.}
    \label{fig:doc-ex}
\end{figure*}

This section showcases an example of a real-world financial document that is used in this work (shown in Figure~\ref{fig:doc-ex}).

\subsection{Document Pre-processor Extractions}
This section showcases examples of tabular and graphical data extracted using the document pre-processor (see Table~\ref{tab:visual1} and Table~\ref{tab:visual2}).

\subsection{GPT Failure Case}
\label{sec:gpt-fail}
This section showcases examples where GPT-$4_o$ fails to accurately interpret chart images (see Table~\ref{tab:gpt-failure}).

\subsection{ChartVLM Failure Case}
\label{sec:chartvlm-fail}
This section showcases examples where ChartVLM fails to completely read and interpret chart images (see Table~\ref{tab:chartvlm-failure}).
Aside from the incorrect value mappings, we also observe that the model entirely overlooked the quarter `Q2/2024' bar data.

\section{Cost Calculation for Popular Commercial Solutions}

Based on analysis of the datasets, a document page contains approximately $600$ tokens. The cost per page of the proposed pre-processing pipeline and commercial solutions are calculated as follows:

\paragraph{LlamaParse:} LlamaParse charges users by `credits'.\footnote{\url{https://docs.cloud.llamaindex.ai/pricing}} The cost is $\$1.0$ per 1000 credits, thus to ensure a fair comparison, GPT-$4_o$ is chosen as the extraction model which costs $30$ credits per page, amounting to $(1/1000) * 30 = 0.03$ per page.

\paragraph{Vertex AI:} Vertex AI's solution\footnote{\url{https://cloud.google.com/vertex-ai/generative-ai/pricing\#token-based-pricing}} employs Gemini-2.0-flash. It can directly process images and charges $\$0.0001935$ token based on the image modality, amounting to $\$0.0001935 * 600 = \$ 0.1161$ per page.

\paragraph{Anthropic:} Anthropic's solution\footnote{\url{https://cloud.google.com/vertex-ai/generative-ai/pricing\#modality-based-pricing}} leverages Claude 3 Opus, which charges $\$15$ per 1M tokens, amounting to $(\$15/1,000,000) * 600 = \$0.009$ per page.

\paragraph{The proposed solution:} Azure Document Intelligence\footnote{\url{https://azure.microsoft.com/en-us/pricing/details/ai-document-intelligence/}} is utilized to extract multi-structured data contents from documents in the pre-processor. It costs $\$375 / 500,000$ pages. Regarding the Azure AI Search index used to store the extracted contents, Microsoft provides a free-tier service, thus the index hosting cost is $\$0$. There are also numerous open-source indexes that can be used instead. Lastly, the model used for generating vector embeddings, Azure OpenAI's ada-embeddings-002, charges $\$0.00006$ page. Azure OpenAI's GPT-$4_o$ charges $\$2.5$ per 1M tokens,\footnote{\url{https://azure.microsoft.com/en-us/pricing/details/cognitive-services/openai-service/}} amounting to $\$0.0015$ per page.
Therefore, the proposed pre-processor used in this work costs $\$0.00075 + \$ 0.0015 + \$0.00006 = \$0.00231$.

\begin{table*}[!bth]
  \begin{tabular}{l} 
    \toprule
     \includegraphics[width=\textwidth]{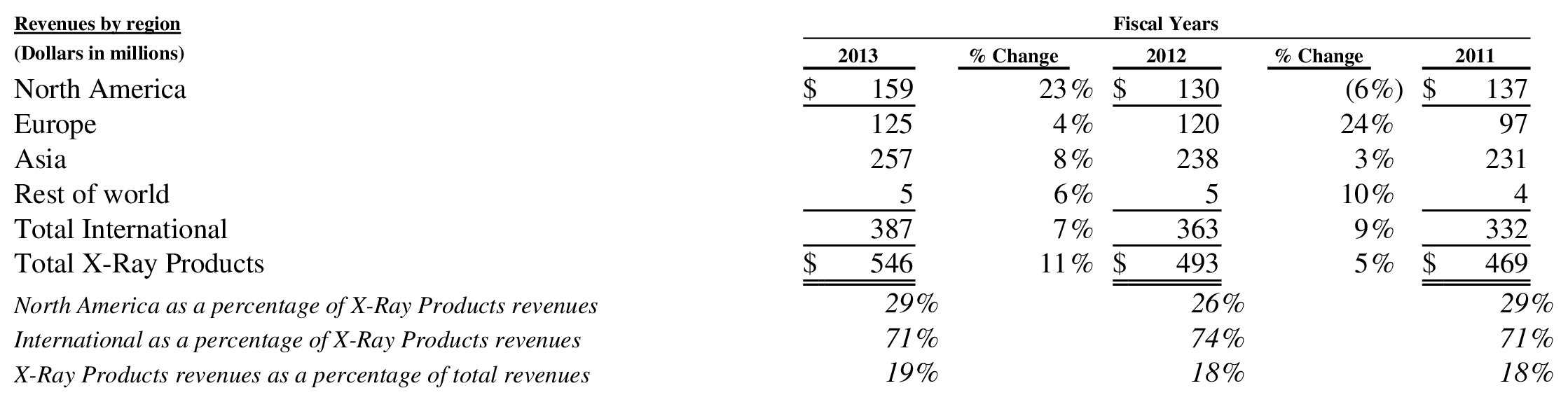}\\
    \midrule
    \midrule
    \hspace{0.5\columnwidth}\texttt{[[\{`Revenues by region;(Dollars in millions);':} \\
    \hspace{0.5\columnwidth}\texttt{`North America',} \\
    \hspace{0.5\columnwidth}\texttt{`Fiscal Years;2013;': `\$ 159',} \\
    \hspace{0.5\columnwidth}\texttt{`Fiscal Years;\% Change;': `(6\%)',} \\
    \hspace{0.5\columnwidth}\texttt{`Fiscal Years;2012;': `\$ 130',} \\
    \hspace{0.5\columnwidth}\texttt{`Fiscal Years;2011;': `\$ 137'\}, ...]]} \\
    \bottomrule
  \end{tabular}
\caption{Visualization for pre-processor table extraction. This is a sampled image from the VQAonBD validation dataset.}
\label{tab:visual1}
\end{table*}
\begin{table*}[bth]
\resizebox{\textwidth}{!}{%
  \begin{tabular}{l} 
    \toprule
     \hspace{0.35\columnwidth}\includegraphics[width=0.8\textwidth,height=10cm]{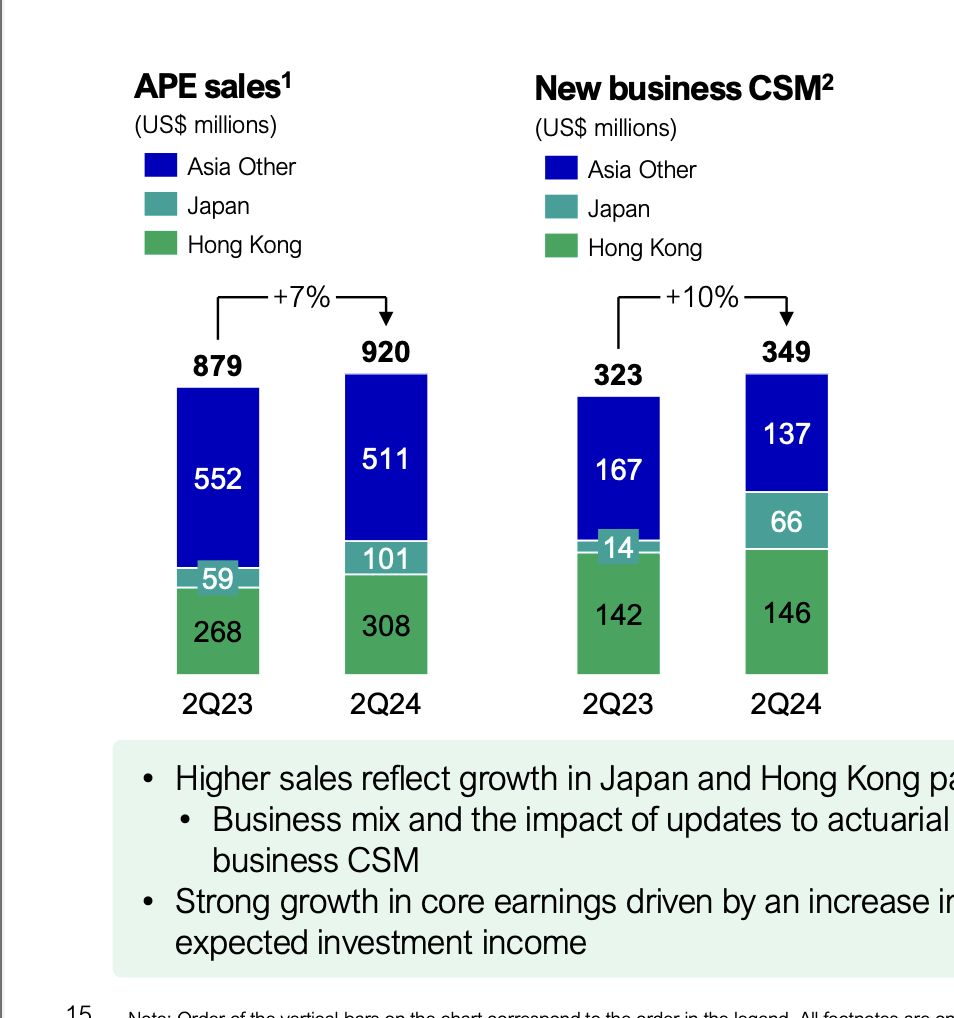}\hspace{0.25\columnwidth}\\
    \midrule
    \midrule
    \hspace{0.7\columnwidth}\texttt{[\{``Quarter ": `` 2Q23 ",}\hspace{0.7\columnwidth}\\
    \hspace{0.7\columnwidth}\texttt{`` APE Sales: Asia other ": 552,}\hspace{0.7\columnwidth} \\
    \hspace{0.7\columnwidth}\texttt{`` APE Sales: Japan ": 59,}\hspace{0.7\columnwidth} \\
    \hspace{0.7\columnwidth}\texttt{`` APE Sales: Hong Kong ": 268,}\hspace{0.7\columnwidth} \\
    \hspace{0.7\columnwidth}\texttt{`` New business CSM: Asia other ": 167,}\hspace{0.7\columnwidth} \\
    \hspace{0.7\columnwidth}\texttt{`` New business CSM: Japan ": 14,}\hspace{0.7\columnwidth} \\
    \hspace{0.7\columnwidth}\texttt{`` New business CSM: Hong Kong ": 142\}}\hspace{0.7\columnwidth} \\
    \bottomrule
  \end{tabular}
 }
\caption{Visualization for pre-processor chart extraction. This is a sampled image from the ComplexChartQA dataset.}
\vspace{-0.5cm}
\label{tab:visual2}
\end{table*}
\begin{table*}[tbh]
  \begin{tabular}{l|l} 
    \toprule
     & \includegraphics[width=0.9\textwidth]{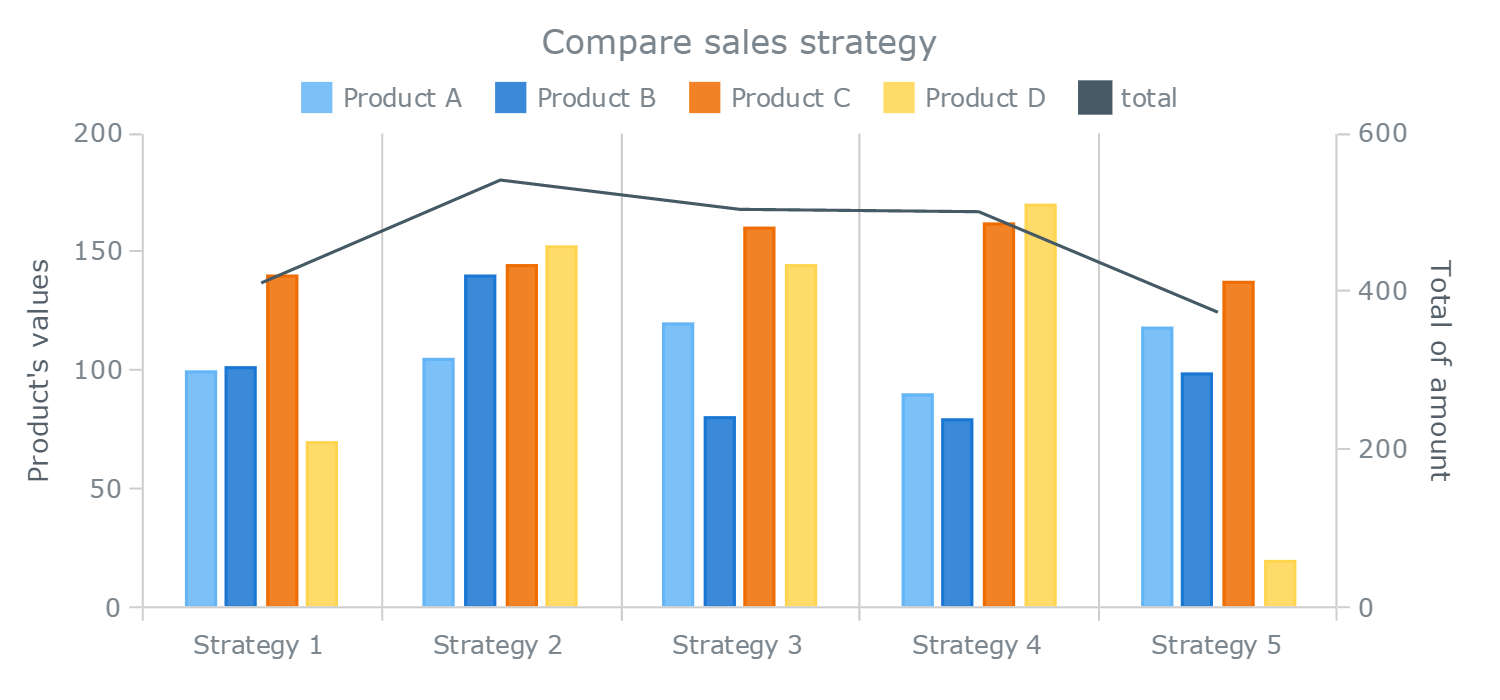}\\
    \midrule
    \midrule
    Question & What is the value of product B value for strategy 5?\\
    Answer & In the chart, the value of Product B for Strategy 5 is 90. \\
    \midrule
    Question & What is the total value for strategy 5?\\
    Answer & The total value for Strategy 5 is 215. \\
    \bottomrule
  \end{tabular}
\caption{Failure case for GPT-4o.}
\label{tab:gpt-failure}
\end{table*}
\begin{table*}[bth]
\resizebox{\textwidth}{!}{%
  \begin{tabular}{l} 
    \toprule
     \hspace{0.35\columnwidth}\includegraphics[width=0.8\textwidth]{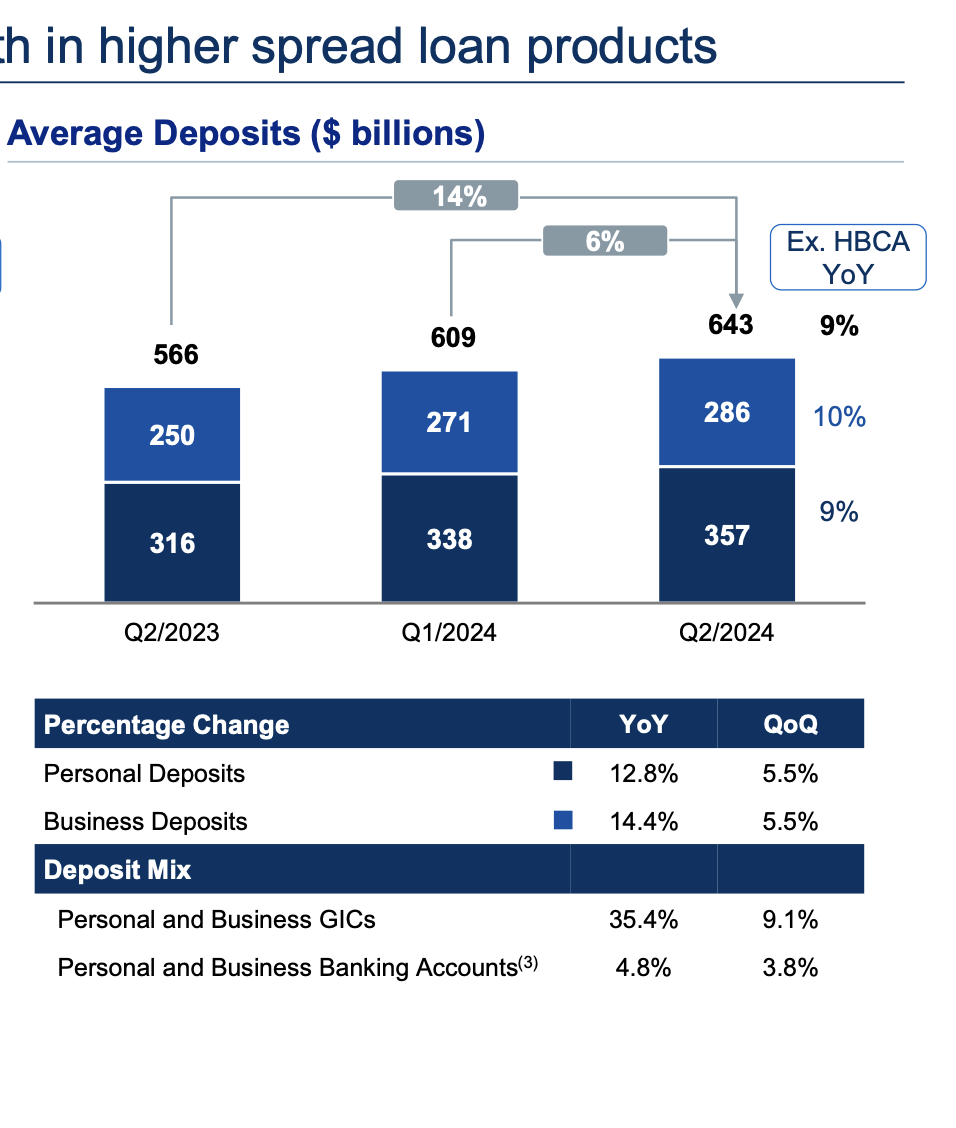}\hspace{0.25\columnwidth}\\
    \midrule
    \midrule
    \hspace{0.7\columnwidth}
    \texttt{[\{``Quarter ": `` Q2/2023 ",}\hspace{0.7\columnwidth}\\
    \hspace{0.7\columnwidth}
    \texttt{`` Average Deposits ": 316,}\hspace{0.7\columnwidth} 
    \\\hspace{0.7\columnwidth}\texttt{`` Percentage Change ": 250, ...\},}
    \\
    \hspace{0.7\columnwidth}\texttt{\{``Quarter ": `` Q1/2024 ",}\hspace{0.7\columnwidth}\\
    \hspace{0.7\columnwidth}\texttt{`` Average Deposits ": 338,}\hspace{0.7\columnwidth} \\
    \hspace{0.7\columnwidth}\texttt{`` Percentage Change ": 271, ...\}]}
    \hspace{0.7\columnwidth} 
    \\
    \bottomrule
  \end{tabular}
 }
\caption{ChartVLM failure case where it overlooks the `Q2/2024' bar data completely due to its limited interpretational capabilities.}
\vspace{-0.5cm}
\label{tab:chartvlm-failure}
\end{table*}
\end{appendices}

\end{document}